\documentclass[traditabstract,a4paper]{aa}
\usepackage{graphicx}
\usepackage{url} 
\usepackage{listings}
\newcommand{\MSun}{\mbox{${\rm M}_\odot$}}

\newcommand{\RSun}{\mbox{${\rm R}_\odot$}}

\def\lteq{\ {\raise-.5ex\hbox{$\buildrel<\over-$}}\ }
\def\apgt{\ {\raise-.5ex\hbox{$\buildrel>\over\sim$}}\ }
\def\aplt{\ {\raise-.5ex\hbox{$\buildrel<\over\sim$}}\ }
\def\lt{\ {\raise-.5ex\hbox{$\buildrel>$}}\ }
\def\gt{\ {\raise-.5ex\hbox{$\buildrel<$}}\ }
\def\eqgt{\ {\raise-.5ex\hbox{$\buildrel>\over-$}}\ }
\def\eqlt{\ {\raise-.5ex\hbox{$\buildrel<\over-$}}\ }

\newfont{\Giga}{cmssbx10 scaled 5200}
\newfont{\giga}{cmssbx10 scaled 4300}
\newfont{\Mega}{cmssbx10 scaled 3200}
\newfont{\mega}{cmssbx10 scaled 2500}
\newfont{\Kilo}{cmssbx10 scaled 2000}
\newfont{\kilo}{cmssbx10 scaled 1600}
\newfont{\Deca}{cmssbx10 scaled 1450}
\newfont{\deca}{cmssbx10 scaled 1200}
\newfont{\Dezi}{cmssbx10 scaled 1100}
\newfont{\dezi}{cmssbx10 scaled 1050}
\newfont{\iGiga}{cmssi10 scaled 6200}
\newfont{\igiga}{cmssi10 scaled 4300}
\newfont{\iMega}{cmssi10 scaled 3200}
\newfont{\imega}{cmssi10 scaled 2500}
\newfont{\iKilo}{cmssi10 scaled 2000}
\newfont{\ikilo}{cmssi10 scaled 1500}
\newfont{\mathGiga}{cmsy10 scaled 6200}
\newfont{\mathgiga}{cmsy10 scaled 4300}
\newfont{\mathMega}{cmsy10 scaled 3200}
\newfont{\mathmega}{cmsy10 scaled 2500}
\newfont{\mathKilo}{cmsy10 scaled 2000}
\newfont{\mathkilo}{cmsy10 scaled 1500}
\newfont{\mathDeca}{cmsy10 scaled 1450}
\newfont{\mathdeca}{cmsy10 scaled 1200}

\def\aap{\ {A\&A}\ }

\def\aj{\ {AJ}\ }

\def\apj{\ {ApJ}\ }
\def\apjl{\ {ApJL}\ }
\def\apjs{\ {ApJS}\ }

\def\mnras{\ {MNRAS}\ }
\def\nat{\ {Nat}\ }

\def\apgt{\ {\raise-.5ex\hbox{$\buildrel>\over\sim$}}\ }
\def\aplt{\ {\raise-.5ex\hbox{$\buildrel<\over\sim$}}\ }
\def\lteq{\ {\raise-.5ex\hbox{$\buildrel<\over-$}}\ }

\begin{document} 

\title{The origin of the two populations of blue stragglers in M30}

\author{
  S.\, Portegies Zwart\inst{1}
}
\offprints{S. Portegies Zwart}
\mail{spz@strw.leidenuniv.nl}
\institute{
 $^1$Leiden Observatory, Leiden University, PO Box 9513, 2300
   RA, Leiden, The Netherlands
}

\date{Received / Accepted }
\titlerunning{Supernova Near the Solar System}
\authorrunning{Portegies Zwart et al.}

\abstract{ We analyze the position of the two populations of blue
  stragglers in the globular cluster M30 in the Hertzsprung-Russell
  diagram. Both populations of blue stragglers are brighter than the
  cluster's turn-off, but one population (the blue blue-stragglers)
  align along the zero-age main-sequence whereas the (red) population
  is elevated in brightness (or colour) by $\sim 0.75$\,mag.  Based on
  stellar evolution and merger simulations we argue that the red
  population, which composes about 40\% of the blue stragglers in
  M\,30, is formed at a constant rate of $\sim 2.8$ blue stragglers
  per Gyr over the last $\sim 10$\,Gyr.  The blue population is formed
  in a burst that started $\sim 3.2$\,Gyr ago at a peak rate of $30$
  blue stragglers per Gyr$^{-1}$ with an e-folding time scale of
  $0.93$\,Gyr.  We speculate that the burst resulted from the core
  collapse of the cluster at an age of about 9.8\,Gyr, whereas the
  constantly formed population is the result of mass transfer and
  mergers through binary evolution.  In that case about half the
  binaries in the cluster effectively result in a blue straggler.

}

\maketitle

\section{Introduction} \label{sec:introduction}

The population of blue stragglers \citep{1953AJ.....58...61S} in M30
appear to be split into two distinct populations
\citep{2009Natur.462.1028F}.  Both populations are brighter than the
current cluster turn-off point in the Hertzsprung-Russel diagram, but
one population that is positioned along the zero-age main sequence
(which \cite{2009Natur.462.1028F} call the blue population) and a
second (red) population that is brighter by about 0.75\, mag.  Both
populations are centrally concentrated.  The majority (90\%) of {\em
  blue} blue stragglers and all {\em red} blue stragglers are within
the projected half-mass radius of the cluster.
\cite{2009Natur.462.1028F} conjecture that the blue population formed
only 1-3 Gyr ago in a relatively short burst triggered by the core
collapse of the cluster.  The red population (60\,\%) has been
attributed to binary mass transfer \citep{2015ApJ...801...67X}, in
which case it should be mainly composed of W~UMa contact binaries
\cite{2017ApJ...849..100J}.

We test these hypotheses by conducting a series of stellar merger
simulations.  We adopt the hypothesis that a blue straggler is the
product of a merger between two stars that merged into a single star
(with mass $M_{\rm tot}$) at some moment in time $t_{\rm mrg}$. Such
a merger can either result from a direct collision during the
dynamical evolution of the star cluster or from an unstable phase of
type A \citep{kw67} mass transfer
\citep{1992ApJ...392..519B,2009Natur.457..288K}, we do not make a
distinction in our models between these two scenarios, as both result
in a single blue straggler.

The moment of merger is determined by finding a merger product that is
consistent with the blue straggler's position in the
Hertzsprung-Russell diagram. From a theoretical perspective we
determine the position of a blue straggler in the Hertzsprung-Russell
diagram by evolving two stars to a certain age $t_{\rm mrg}$, perform
the merger calculation and continue to evolve the merger product for
the remaining age of the cluster \citep[13\,Gyr according
  to][]{1996yCat.7195....0H}. As we explain in the following section,
the position of a blue straggler in the Hertzsprung-Russell diagram is
sensitive to the total mass of the merger product as well as to the
moment of the merger, but rather insensitive to the individual masses
of the two stars at birth.

\section{The experimental setup} \label{sec:experiment}

We adopted the {\tt MESA} Henyey stellar evolution code
\citep{2011ApJS..192....3P} to model the evolution of the stars with
$[Fe/H] =-2.33$ \citep[which according to their Tab.\,1 is consistent
  with the cluster's metallicity][]{2009A&A...508..695C}.
%% was $[Fe/H] =-2.33$
Both stars are initialized at the zero-age main-sequence
and evolved to $t_{\rm mrg}$. At that moment we merge the two stars
using Make-Me-A-Massive-star \citep{2008MNRAS.383L...5G}, which uses
Archimedes' principle to calculate the structure of the star resulting
from a merger between two stars. After this we continue to evolve the
merger product using {\tt MESA} to the age of the cluster M30.

The numerical setup is realized with the Astronomical Multipurpose
Software Environment
\citep[AMUSE,][]{portegies_zwart_simon_2018_1443252,2013CoPhC.183..456P,2013AA...557A..84P}.
Our analysis is comparable to the method described in
\cite{2002ApJ...568..939L}, but then our procedure is completely
automated.  We tentatively limit ourselves to head-on collisions, such
a described in \cite{1997ApJ...487..290S} because off-centre
collisions do, except from some additional mass loss, not seem to
result in qualitative differences in the merger product \citep[][see
  also Chapter 5.3.3 of \cite{AMUSE}]{2001ApJ...548..323S}.

We initialize a grid of primary masses between 0.5\,\MSun\, and the
turn-off mass of 0.85\,\MSun\, in steps of 0.05\,\MSun\, and secondary
masses between 0.2\,\MSun\, with the same upper limit in steps of
0.005\,\MSun.  The merger time is chosen between 0.1\,Gyr and the
age of the cluster with steps of 0.98\,Gyr.  The evolutionary state of
the merger product at any time after the collision is predominantly
determined by the total mass of the merger product $M_{\rm tot}$.
Small variations in the mass lost during the collision therefore have
little effect on our determination of the merger time, because the
location in the Hertzsprung-Russell diagram then depends on the total
mass of the merger product and the moment of collision, rather than on
the masses of the two stars that participate in the merger.

In Appendix A we present the {\tt AMUSE} script to reproduce the
calculations in this paper.

\section{Results} \label{sec:Results}

\begin{figure}
\centering
\includegraphics[width=0.95\columnwidth]{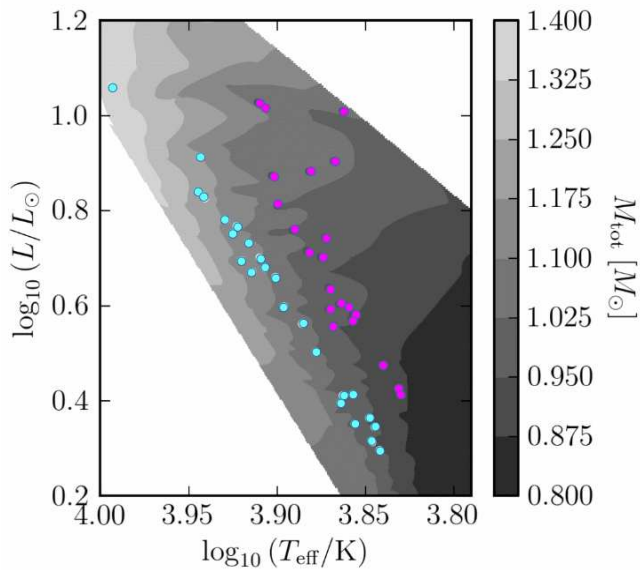}
\caption{Hertzsprung-Russell diagram of the M30 blue stragglers. The
  original data is from \cite{Ferraro2015} was convoluted to the
  temperature-luminosity plane.  With effective temperature and
  luminosity from \cite{2009Natur.462.1028F}.  The {\em blue} and {\em
    red} blue stragglers are indicated as such.
\label{fig:CM_masstot}}
\end{figure}

The Hertzsprung-Russell diagram of the blue stragglers is presented in
Fig.\,\ref{fig:CM_masstot}.  Overplotted in colour, is the total
mass of the merger products that remain on the main-sequence until
an age of 13\,Gyr.

Information about the masses of the two stars is largely lost in the
merger process, and can hardly be used for diagnostics \cite[see
  also][]{2002ApJ...568..939L}. We, therefore, use the total blue
straggler mass and the merger time as a diagnostic tool.

\begin{figure}
\centering
\includegraphics[width=0.95\columnwidth]{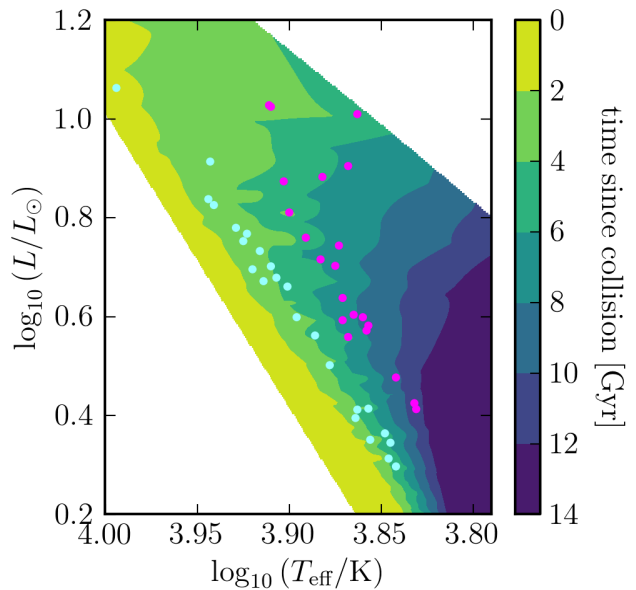}
\caption{Same as Fig.\,\ref{fig:CM_masstot}, except for the time since
  collision, which is color coded here.
\label{fig:CM_tcoll}}
\end{figure}

\begin{figure}
\centering
\includegraphics[width=0.95\columnwidth]{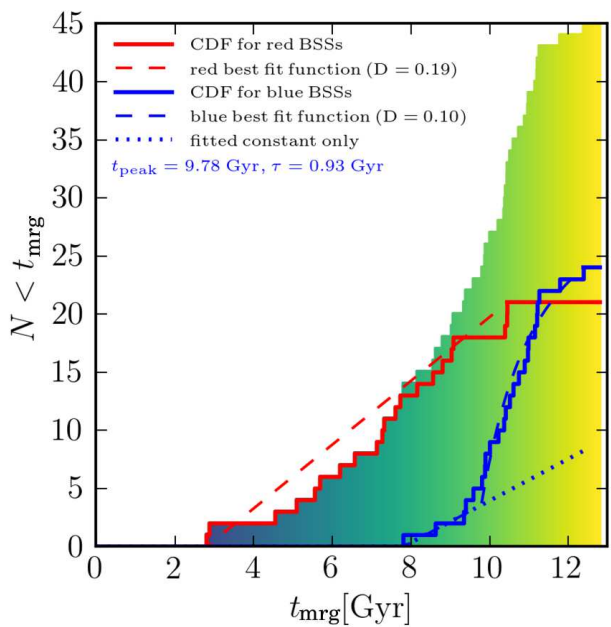}
\caption{The cumulative distribution of merger times ($t_{\rm mrg}$)
  for all the blue stragglers (blue plus red as the colour shaded area
  where the color corresponds to that in
  Fig.\,\ref{fig:CM_tcoll}). The solid blue and solid red curves give
  the cumulative distribution for the blue and {\em red} blue
  stragglers, respectively.  The dashed and dotted blue curves give
  the fit to the {\em blue} blue stragglers (the dotted curve gives
  the linear component and dashes give the sum of the exponential and
  the linear fits).  The red dashed curve gives the linear fit to the
  {\em red} blue stragglers. The color coding is identical to that
  used in Fig.\,\ref{fig:CM_tcoll}, here indicating the time since the
  merger occurred.
\label{fig:cdf_tcoll_blue_and_red}}
\end{figure}

In fig.\,\ref{fig:CM_tcoll} we present the same data as in
fig.\,\ref{fig:CM_masstot}, but now overplotted in color is the time
since collision. The lightest shades indicate the most recent mergers.
The {\em blue} blue stragglers tend to cluster around a time since
collision between 1\,Gyr and 3\,Gyr ago (in light-green), whereas the
{\em red} blue stragglers span a much wider range of merger times.  We
quantify this statement in fig.\,\ref{fig:cdf_tcoll_blue_and_red},
where we present the cumulative distribution of merger times for the
blue and red blue-stragglers together (colours) and separately (solid
curves).

We fitted both distributions with a constant blue straggler formation
rate combined with a burst and exponential decay.  The best fits are
obtained using the Nelder-Mead simplex optimization
\citep{1965TCJ...007..004N} to find the minimum Kolmogorov--Smirnov
(KS) statistic over the free parameters $t_{\rm mrg}$, and
e-folding time scale $\tau$, in combination with a line describing
the constant formation rate.

The best fit (with KS statistics $D=0.10$, $p=0.24$) to the {\em blue}
blue stragglers is obtained for $t_{\rm mrg} = 9.8$\,Gyr, $\tau =
0.93$\,Gyr with a peak formation rate of 30 blue stragglers per Gyr
and an additional constant formation rate of $1.8\pm0.6$ per Gyr.

Fitting the {\em red} blue straggler formation rate with the same set
of functions (a constant rate plus a power-law) did not result in a
satisfactory fit, but a single linear formation rate did produce the
KS statistic of $D=0.19$ ($p=0.23$) with a constant formation rate of
only $2.8\pm0.5$ per Gyr between an age of 3\,Gyr to 10\,Gyr.  It is
interesting to note that the formation rate for the red population
levels off when the blue population reaches its maximum rate.

\section{Interpretation}

The majority of blue stragglers in star clusters are thought to
originate from either stellar collisions \citep{1989AJ.....98..217L}
or from mass transfer in a close binary system
\citep{1984MNRAS.211..391C}. We will argue here that the two distinct
populations found in M30 can be attributed to these different
formation channels (see also \cite{2009Natur.462.1028F}).  We argue
that the red population is consistent with being formed continuously
and through mass transfer and mergers in binary systems, whereas the
blue population is mainly the result of collisions during the core
collapse of the star cluster. In that perspective, we attribute the
burst population to the collision scenario, whereas the continuously
formed population is the result of binary evolution.

\subsection{The burst population of blue stragglers}

According to our analysis about one third (15) of the blue stragglers
in M30 are formed in a rather short burst that started at 9.8\,Gyr
with power-law decay with a characteristic time scale of 0.9\,Gyr. At
the peak the blue stragglers in the burst formed at a rate of about
30\, blue stragglers per Gyr. But due to the exponential, we adopted
(and satisfactorily fitted) this burst lasts only a short while, long
enough to produce some 20 blue stragglers.

We estimate the expected formation rate through stellar mergers during
core collapse. This is realized by calculating the collision rate
\begin{equation}
  \Gamma_{\rm mrg} = n \sigma v.
\end{equation}
Here $n$ is the stellar number density, $v$ the velocity dispersion,
and the approximate gravitational-focused cross-section $\sigma$ is
\begin{equation}
  \sigma  = r \nu^2.
\end{equation}
Here $\nu \equiv v/v_\infty$ is the stellar velocity dispersion as
fraction of the stellar escape speed \citep{1987gady.book.....B}.

\cite{2004MNRAS.349..129D} derived a formation rate of blue stragglers
for a star cluster through direct stellar collisions, using the above
arguments. We can adopt their eq.\,4 to calculate the expected number
of blue stragglers formed through collisions.  By adopting the current
observed cluster parameters ($n \simeq 3.8\cdot 10^5 {\rm pc}^{-3}$,
$N=1.6 \cdot 10^5$ stars, $r_{\rm core} \simeq 0.2$\,pc and adopting a
mean stellar mass of 0.5\,\MSun\, from \cite{1996yCat.7195....0H}) we
arrive at the current average blue-straggler production-rate through
collisions of 20\,Gyr$^{-1}$.

\subsection{The continuously formed blue stragglers}

Mass transfer in binary systems are less likely to depend strongly on
the cluster core density because binaries are present in the halo as
well as in the cluster centre, which causes them to be more
homogeneously distributed across the cluster
\citep{1992ApJ...389..527H}, whereas direct stellar collisions are
predominantly occurring at the very centre of the cluster
\cite{1997A&A...328..130P}.  The binary merger rate is also not
expected to be particularly affected by the cluster density
profile. We, therefore, argue that the constant rate is a result of
binary mass transfer and coalescence.

We can constrain the underlying binary semi-major axis distribution
and mass ratio distribution that produces a constant blue straggler
formation rate (or a constant rate of binaries that engage in a phase
of mass transfer).  Mass transfer in a binary system is typically
initiated by the primary star, which overfills its Roche lobe when it
either ascend the giant branch or, for very tight binaries, along the
main sequence. Since the timescale between the terminal-age
main-sequence and the post-AGB phase is only a small fraction ($\aplt
0.15$) of the main-sequence lifetime, we adopt the main-sequence
lifetime as the limiting factor between zero age and the start of
Roche-lobe overflow.

The lifetime of a main-sequence star scales as $t_{\rm ms} \propto m^{2.5}$\,
\cite{1962pfig.book.....S}.  A primary mass distribution of $f(m)
\propto m^{-2.35}$ \cite{1955ApJ...121..161S} then produces a roughly
constant rate at which stars leave the main sequence, consistent with
the observed constant rate of blue-straggler formation.

M30 has a binary fraction of about 3\%
\citep{1991ASPC...13..443R,2012A&A...540A..16M}, so with $1.6 \cdot
10^5$ stars the cluster has 4800 binaries.  A standard Salpeter mass
function has about 5.8\% of the stars between 0.5\,\MSun\, and $\sim
0.85$\,\MSun. A 0.5\,\MSun\, star requires an equal mass secondary
star to evolve into a blue straggler in an unstable phase of mass
transfer, whereas a 0.85\,\MSun\, star only requires a companion with
a mass of $\apgt 0.1$\,\MSun\, \citep{1997A&A...328..130P}. On average
about half the binaries in the appropriate mass range then produce
blue stragglers, totalling a potential number of 280. Because the
mass-ratio distribution in cluster binaries tends to be flat
\cite{2007A&A...474...77K}, roughly half of these binaries have a
total mass that upon a merger results in a blue straggler. The
binaries with small mass ratio do not form as a blue straggler
directly upon the merger because the total mass of the merger product
does not exceed the turn-off mass, but these stars pop-up later when
their rejuvenation causes them to stay behind in their evolution
\citep{1997A&A...328..143P}.  The orbital separations of primordial
binaries range from a few $\RSun$ and a maximum of $\sim 10^4$\,AU at
the \cite{1975MNRAS.173..729H} limit for hard-soft binaries.
Roche-lobe overflow on the main-sequence is most favourable for the
formation of blue straggler. This process is effective for binaries
with an orbital separation of $\aplt 10$\,\RSun. With a flat
distribution in the logarithm of the semi-major axis
\citep{2004RMxAC..21...33Z,2007A&A...474...77K}, only about one in
four binaries will be effectively producing a blue straggler
\citep{2009MNRAS.395.1822C}. The entire binary reservoir then produces
$\sim 35$ blue stragglers through mass transfer or coalescence.

\section{Discussion} \label{sec:Discussion}

The $13~\mathrm{Gyr}$--old globular cluster M30 has a rich population
of blue stragglers, which appear to be distributed bimodally in the
Hertzsprung-Russell diagram \citep{2009Natur.462.1028F}.  We tested
the hypothesis that all these blue stragglers are the result of a
merger due to an unstable phase of mass transfer in a binary system or
a stellar collision.  We simulate the current population of blue
stragglers that could have resulted from the coalescence of two stars
at some time $t_{\rm mrg}$ with a total mass of $m_{\rm tot}$. The
stellar merger product was subsequently evolved to the current age of
the cluster of 13\,Gyr.  For each point in the Hertzsprung-Russell
diagram we then obtain a unique solution for the mass of the blue
straggler and the moment of merger $t_{\rm mrg}$. The two masses of
the stars that merge are not well discriminated in the results,
because the memory of the two stellar masses is lost in the merger
process due to the mixing in the merger process \cite[][and much later
  literature]{1987ApJ...323..614B}.

The merger time-distribution for the blue blue stragglers is best
described by a peak of formation of $\sim 30$ blue stragglers per Gyr
at $t_{\rm mrg} \simeq 9.8$\,Gyr and an e-folding time scale of
$0.93$\,Gyr superposed with an additional constant formation rate of
$1.8$\, per Gyr between $t_{\rm mrg} \simeq 8$\,Gyr and the age of the
cluster.  This is consistent with the conjecture by
\cite{2009Natur.462.1028F} that these blue stragglers ware born in a
burst during the core-collapse phase of the cluster some 2--3\,Gyr
ago.  The population of red blue stragglers is best described with a
constant formation rate of 2.8\,Gyr$^{-1}$ between an age of $t_{\rm
  mrg} \simeq 3$\,Gyr and 11\,Gyr (between 11 and 2\,Gyr ago).

About 10\% of the blue and red blue stragglers appear to be missed in
the observational data.  We interpret this bimodality of blue
stragglers with two distinct channels through which they form, much in
the same way as \cite{2009Natur.462.1028F} argued based on the
observations.  The continuously formed population is consistent with
originating from mass transfer in primordial binaries. In that case, about 10-15\% ($\sim 35/280$) of any binary leads to the formation of
a blue straggler at a constant rate. The burst with an exponential
decay in the formation of blue stragglers is the result of direct
stellar mergers during the core collapse of the star cluster.

We attribute the start of the blue straggler formation burst to the
moment of core collapse in the star cluster, at an age of $\sim
9.8$\,Gyr.  This is consistent with the inverse cluster-evolution
analysis by \cite[][for the details of the analysis, but the results
  adopted here were presented at the IAU conference in
  2015]{2015MNRAS.453..605P}, which leads to a core collapse at $9.5
\pm 0.4$\,Gyr and which is consistent with the start of the
blue-straggler formation burst.

We conclude that the core collapse of the cluster was associated with
a burst in the formation of blue stragglers. The exponential decay is
a result of the relatively extended period during which the cluster
remains in a collapsed --or post-collapsed-- state following the
primary collapse \citep{1989MNRAS.237..757H}. This could indicate a
prolonged period of gravothermal oscillations following the primary
collapse of the cluster core \citep{1994MNRAS.271..706H}. We argue
that the post-collapsed phase lasted for about $1$\,Gyr.  The current
cluster has a relatively low density consistent with the late stages
of post-collapse \cite{1984MNRAS.208..493B}.

\section*{Acknowledgments}
It is a great pleasure to thank Alex Rimoldi for the preliminary
analysis and making the figures for this paper and Tjibaria Pijloo for
reconstructing the cluster's moment of core collapse.  This work was
in part done at the Canadian Institute for Theoretical Astronomy and I
am grateful for their support, in particular to Norm Murray who made
this possible.

\section*{Appendix: minimal AMUSE script for the runs}

In the listing we present an AMUSE
\citep{AMUSE,portegies_zwart_simon_2018_1443252} script to calculate
the evolution of two stars that underwent a merger, and that is
continued to evolve to some late time. This script is tuned for M30,
to limit the number of input parameters.  The script starts by
initializing the {\tt MESA} stellar evolution code
\citep{2010ascl.soft10083P,2011ApJS..192....3P}, declare the two stars
and submit them to the stellar evolution code.  In the subsequent
block of lines, the stars are evolved to an age of 9\,Gyr, and the
resulting stellar models are merged using {\tt MakeMeAMassiveStar}
\citep{2008MNRAS.383L...5G,2014ascl.soft12010L}.  In the last block of
lines, the merger product is submitted again to the stellar evolution
code and continued to evolve to an age of 13\,Gyr.  For a more
thorough explanation of the script we refer to the AMUSE book
\citep{AMUSE} Chapter 4 (List.\,4.7) or see \url{amusecode.org} for
the complete source.

\begin{lstlisting}[linerange={0-29},float,label=Src:collide_two_stars.py]
import numpy
from amuse.lab import *

def merge_stars(Mprim=0.8|units.MSun,
                Msec=0.6|units.MSun,
                t_coll=11|units.Gyr):

    code = MESA()
    p = Particles(mass=Mprim)
    primary = code.particles.add_particle(p)
    s = Particles(mass=Msec)
    secondary = code.particles.add_particle(s)
    code.evolve_model(t_coll)
    stars = code.particles
    print "Pre merger:\n", stars

    n_zones = stars.get_number_of_zones()
    code.merge_colliding(primary.copy(),
      secondary.copy(),
      MakeMeAMassiveStar,
      dict(),
      dict(target_n_shells_mixing=max(n_zones)),
      return_merge_products=["se"])
    code.evolve_model(13|units.Gyr)
    print "Post merger stellar parameters:",\
      "T=", stars.temperature,\
      "L=", stars.luminosity.in_(units.LSun)

    code.stop()
\end{lstlisting}

\end{document}